\documentclass[doublecol]{epl2} 

\title{Spin inversion devices operating at Fano anti-resonances}
\shorttitle{Spin inversion devices operating at Fano anti-resonances} 

\author{J. L. Cardoso \and P. Pereyra}
\shortauthor{J. L. Cardoso \and P. Pereyra}

\institute{                    
  \inst{1} F\'{\i}sica Te\'orica y Materia Condensada, UAM-Azcapotzalco,
Avenida San Pablo 180, C\'odigo Postal O2200 M\'exico Distrito Federal,
M\'exico
}

\abstract{Using the exact two-propagating-modes solutions for electrons in quasi-2D semiconductor wave-guide, under sectionally constant magnetic fields and spin-orbit interactions, it is explicitly shown that the Fano-like resonances and antiresonances lead to sudden suppressions and enhancements of $T_{\uparrow,\uparrow}$ and $T_{\downarrow,\downarrow}$, as well as of the spin-transition probability $T_{\downarrow,\uparrow}$. Our calculations show that when the magnetic-field-tilting angle $\theta_H$ is increased, the spin-field interaction becomes the most significant mechanism for the spin transitions in magnetic superlattices. Taking advantage of these spin-transport effects, simple and efficient spin-inverter devices are proposed. To better visualize the relative influence of the specific semiconductor properties on the device performance and device characteristics, we consider two magnetic superlattices based on semiconductors having entirely different Land\`e g factor: $GaAs$ and $InSb$. Although slightly more efficient spin-inversion devices are obtained for the $GaAs$ semiconductor, $InSb$ requires lower magnetic fields and its efficiency can be close to $80\%$.}

\begin{document}

\maketitle
The design of simple and efficient devices working as reliable sources of spin-polarized electrons\cite{IZutic} is one of the main goals of the spintronics field. This purpose is still far from being fully achieved, both theoretically and experimentally. On the experimental side, spin-injecting devices into semiconductor structures have evolved from ferromagnetic metallic contacts \cite{PRHammar, GAPrinz} and diluted magnetic semiconductor spin-aligner structures \cite{Fiederling, Ohno, TKoga, MKohda, EJohnston}, to optical systems sensitive to spin orientation \cite{YKato, JMKikkawa1, JMKikkawa2}. On the theoretical side, the spin carriers transport problem spurred an intense research activity and different kind of approaches were applied to an equally large diversity of systems, from open quantum dots \cite{JFSong, METorio} and magnetic multilayers \cite{Datta,Barthelemy,Johnson,Berger} to magnetic superlattices\cite{JLCardoso, Wu}. Concerning the control of spin-carrier currents, one of the relevant problems is the manipulation at will of the spin orientation, i.e. the ability to polarize, unpolarize and invert the carriers spin orientation whenever it is necessary. In this letter we discuss physical conditions for an efficient spin-inverter device. Among the numerous spin-filter devices proposed in the literature, it has been shown that the homogeneous magnetic superlattice (HMSL) can, in principle, work as a $100{\%}$ spin-polarizer \cite{JLCardoso} and as a simple device to induce (by the spin-orbit interaction and by the $x-y$ field components \cite{JLCardoso2}) spin-flipping processes. These processes, visualized through the spatial evolution of the spin-carrier wave functions\cite{JLCardoso2}, lead us to suggest here physical conditions for a simple and highly efficient two-step spin-inverter device, with a HMSL of length $L$ for each step.

\begin{figure}
\onefigure[angle=0,width=80mm]{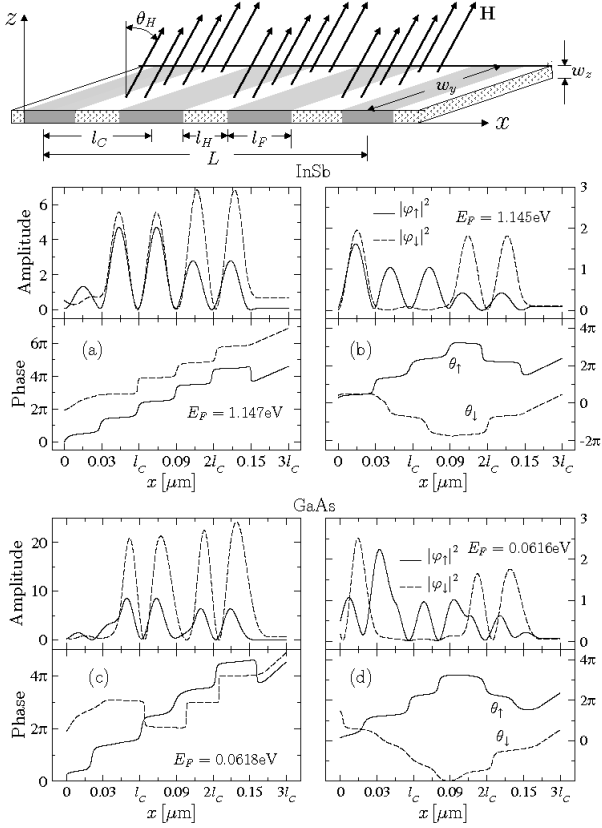}
\caption{Distribution of the electron-spin polarization along a magnetic superlattice. In the upper part, the HMSL is shown as a quasi-2D semiconductor wave guide subject to an external homogeneous magnetic field $\mathbf H$ acting only on alternating stripes of length $l_H$. In the lower graphs we plot the wave-functions (amplitudes and phases) for InSb ($g=50.6$) and GaAs ($g=0.41$), assuming that the incident electron beam is unpolarized. In (a) and (c) the energies are chosen such that $T_{\downarrow, \uparrow}$ is larger than in (b) and (d), with $T_{\downarrow, \downarrow}\gg T_{\uparrow, \uparrow}$ (see specific values in the text). In the l.h.s. graphs the spin $\uparrow$ function looses its identity and follows the behavior of $\varphi_{\downarrow}$. They are practically in phase for $x\geq l_C/2$, On the r.h.s. $\varphi_{\uparrow}$ and $\varphi_{\downarrow}$ remain independent. For this to occur, we need $T_{\downarrow, \uparrow}\simeq 0$ and $T_{\downarrow, \downarrow}\simeq T_{\uparrow, \uparrow}$.}
\label{superl}
\end{figure}

If we have a quasi-2D semiconductor waveguide of length $L_S>2L$ and transversal widths $w_y$ and $w_z$ (with $w_z \ll w_y$), one can produce two consecutive magnetic superlattices by coating alternating stripes of length $L_F$ with some field-screening material\footnote{We assume that a partial or total screening of the magnetic field is possible using $\mu$-metals, superconductor paintings, etc.}. Thus, an external magnetic field $\bf{H}$, acts only on the uncoated stripes of length $L_H$ and a magnetic superlattice $FHFHFHF \dots$ is formed along the $x$ axis (see fig. \ref{superl}). The HMSL has the great advantage that it makes unnecessary to grow layers of different types of materials or doping with magnetic impurities. With a HMSL one can have, at reduced cost, all the desired spin-field and spin-orbit interactions, together with the highly relevant coherence phenomena. Though the HMSL is in some respects similar to the widely studied diluted magnetic structures, it possesses other important advantages and differences that make it not only a more feasible system to produce, but also a system where the electron injection to and from the device is simpler. The most attractive feature of these systems rests on its analytical simplicity tractability. In fact, we have been able to completely solve, in the lowest two propagating modes approximation, the Shr\"{o}dinger equation in the presence of the Dresselhaus and the Rashba spin-orbit interactions. Therefore, the spin dynamics is fully described within this approximation. Though some basic quantities like the magnetic field strength $H$, the energy levels separation $\Delta E_{\nu\nu'}$, the subbands separation $\Delta E_{\mu\mu'}$ (related in some way to the operational temperatures $T$) and the geometrical parameters are deeply intertwined, the competition between the spin-field and the spin-orbit Rashba and Dresselhass interactions is fundamental, particularly in device behaving as the one discussed here. To illustrate the application of our solutions\footnote{Details will be published elsewhere} we present here calculations to visualize the evolution of the spin-polarization phenomena along the superlattices and for the spin inverters assuming that the Rashba coefficient can be neglected. This calculations are performed using $GaAs$ and $InSb$ magnetic superlattices having very different Land\`{e} $g$-factors, and Dresselhaus coefficients that differ by 1 order of magnitude. $InSb$ magnetic superlattices with $L_H\sim 10nm$ subbands separation of $\sim 15$meV require external magnetic fields of the order of $0.25$T, while $GaAs$,also with $L_H\sim 10nm$ and $\Delta E_{\mu\mu'}\simeq 15$meV, require fields of $\sim 1.5$T. To increase the subband separation, we need larger fields and/or smaller $L_H$'s. It is clear that one can expect that by increasing the subbands separation the operational temperature $T$ will also increase. However, this may not be enough. We expect that the physics explained here will apply at low temperatures.

In our approach\cite{JLCardoso}, the wave function $\psi(x)=\varphi_{\uparrow}(x)+\varphi_{\downarrow}(x)$ in a semiconductor region under an external magnetic field $ {\bf{H}} = H \left( 0, \tan \theta_{H}, 1 \right) $, is written as
\begin{eqnarray} \label{transm-H}
\left ( \begin{array}{c} \varphi_{\uparrow}(x) \\ \varphi_{\downarrow}(x) \end{array}\right )= \left[ (1+ik){\bf
A}(x)+{\bf r}_n (1-ik){\bf B}(x)\right] \left(
\begin{array}{c}
a_{\uparrow} \\
a_{\downarrow}
\end{array} \right),
\end{eqnarray}
with $a_{\uparrow}$ and $ a_{\downarrow}$ the incoming amplitudes, $k$ the longitudinal Fermi wave number, $ {\bf r}_n $ the $n$-cells reflection amplitude
\begin{equation} \label{refcoeff}
{\bf r}_n = \left( \begin{array}{cc}
r_{n\uparrow, \uparrow} & r_{n\uparrow, \downarrow} \\
r_{n\downarrow, \uparrow} & r_{n\downarrow, \downarrow} \end{array} \right),
\end{equation}
and ${\bf A}(x)$ and  ${\bf B}(x)$ the Hypergeometric matrix functions
\begin{eqnarray}
{\bf A} &=& {}_1 {\bf F}_1 \left( -\frac{ {\bf b}}{2} ; \frac{1}{2}; \frac{  l_H^2 \cos \theta_H }{l_{B}^2}
\right) e^{-  l_H^2 \cos \theta_H /2l_{B}^2},
\\
{\bf B} &=& l_H {}\,_1{\bf F}_1 \left( \frac{{\bf I}-{\bf b}}{2}; \frac{3}{2}; \frac{  l_H^2 \cos \theta_H
}{l_{B}^2} \right) e^{-  l_H^2 \cos \theta_H /2 l_{B}^2}.\,\,\,\,\,\,\,\,\,
\end{eqnarray}
Here $l_B$ is the magnetic length and
\begin{equation} \label{sigma-matrix}
\begin{array}{c}
{\bf b} = {\bf I} \left( \frac{ \kappa^2 l_{B}^2}{2 \cos \theta } - \frac{1}{2} \right) + \frac{g}{4} \left[\,\,
\tan \theta \,\, {\mbox{\boldmath{ $\sigma$}}}_x  - {\bf J}_{1,1}\,\,{\mbox{\boldmath{ $\sigma$}}}_z \,\,\right],
\end{array}
\end{equation}
with {\bf I} the unit matrix, ${\mbox{\boldmath{ $\sigma$}}}_i$ the well-known Pauli matrices and
\begin{eqnarray}
{\bf J}_{1,1} = \frac{\pi^2 \left[\,\,{\bf I}\sin \left( 2\beta w_y \right)+i{\mbox{\boldmath{ $\sigma$}}}_y
\left( \cos \left( 2\beta w_y \right) - 1 \right)\,\,\right]}{2\beta w_y \left( \beta^2 w_y^2 -\pi^2 \right)}
\end{eqnarray}

To calculate the $n$-cells superlattice scattering amplitudes ${\bf r}_n$ and ${\bf t}_n$, we use the well-known relations of the scattering approach to multichannel finite periodic systems\cite{PPereyra1}. As usual, the $i,j$-th element of the transmission coefficient $(T_n)_{i,j}=({\bf t}_n{\bf t^\dagger}_n)_{i,j}$ is taken as the transmission from the incoming channel $j$ on the left to the outgoing channel $i$ on the right. The conductance, resonant energies and wave functions, among other physical quantities appear naturally in this approach. In fact, we use all these quantities as functions of the field strength $H$, the tilting angle $\theta_H$ and the geometrical parameters to visualize the performance of different HMSL's configurations, searching for the HMSL with the desired properties. In the {\it absence of spin-mixing} (when $\beta_D=0$ and $\theta_H=0$) ${\bf r}$ and ${\bf t}$ are diagonal.
\begin{figure}
\onefigure[angle=0,width=80mm]{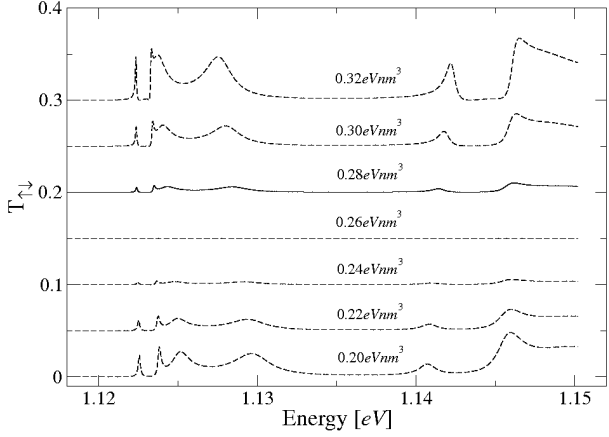}
\caption{The transition coefficient $T_{\uparrow, \downarrow}$ as function of the energy for different values of the Dresselhaus coefficient $\beta$. In these graphs $g=50.1$ the InSb Land\`{e} $g$-factor}
\label{Dresselhaus}
\end{figure}
As mentioned before, clear signatures of coherent spin flipping processes can be inferred from the carrier's wave function behavior. To help visualizing part of the rich phenomenology emerging from our results, we show in the lower panels of fig. \ref{superl} the evolution of spin $\uparrow$ and spin $\downarrow$ wave-function amplitudes (WFA's) and phases, through a simple $3$-cell magnetic superlattice, when the spin-flip transition coefficients $T_{\uparrow, \downarrow}$ are relatively large (left column) or relatively small (right column). By Choosing appropriately the SL geometrical parameters, the field strength and the tilting angle, or by varying slightly the Fermi energy (as was done in the exercise shown in fig. 1), one can fix the transmission and reflection coefficients in such a way that the wave function $\varphi_\uparrow $ looses or keeps its identity as one moves along the propagation direction inside the superlattice. Though the incoming electron beams in (a), (b), (c) and (d) are equally unpolarized, the WFA's inside the SL's are different. In the upper panels (a) and (b) we plot the wave function inside an InSb magnetic SL for $L_F=50$nm, $L_H=10$nm, external magnetic field $H=200$mT and tilting angle $\theta=40^o$. In (a) the Fermi energy is resonant for spin ${\downarrow}$ with $T_{\downarrow, \downarrow}\simeq 0.84$ and $R_{\uparrow, \uparrow}\simeq 0.006$, while $T_{\uparrow, \uparrow}=0.49$ and $R_{\downarrow, \downarrow}\simeq 0.35$. Because of the relatively large {\it transition coefficient} $T_{\uparrow, \downarrow}\simeq 0.11$, the amplitude and phase of $\varphi_\uparrow $ inside the SL tends to follow the behavior of $ \varphi_\downarrow $, and consequently both functions are practically in phase. This means that beyond the first cell the population of the spin $\uparrow$ electrons contains an important fraction of electrons coming from the spin $\downarrow$ beam. In (b) the energy is not anymore resonant with reflection coefficients larger that the transmission ones. $T_{\downarrow, \downarrow}$ and $R_{\downarrow,\downarrow}$ are almost the same as for spin ${\uparrow}$ ($T_{\downarrow, \downarrow}\simeq 0.19$ and $R_{\downarrow, \downarrow}\simeq 0.80$, $T_{\uparrow, \uparrow}\simeq 0.21$, and $R_{\uparrow, \uparrow}\simeq 0.78$), while $T_{\uparrow, \downarrow}=0.002$. Thus, the wave function $\varphi_\downarrow$ features remain independent from those of $\varphi_\uparrow$ all the way along the SL. In fact, from the lower panels in fig. 1, it can be seen that the general trend of the wave functions (amplitude and phase) for $GaAs$ and $InSb$ magnetic superlattices are similar. As mentioned before, the main difference between these systems resides in the magnitude of the magnetic field that has to be applied in order to observe the resonant behavior, a condition that is crucial to induce the spin-flip phenomenon. Our results show that the contributions ratio of the two sources of spin precession, spin-field and the spin-orbit interaction, vary in a complicated way. In fact, as shown in fig. \ref{Dresselhaus} the transition coefficients $T_{\uparrow, \downarrow}$, plotted for different values of the Dresselhaus coefficient $\beta$ and fixed Land\`{e}-factor (in this case $g=50.1$), are highly sensitive to $\beta$, which was varied in the vicinity of the experimental value: $\beta=0.28$eV/nm$^3$. Near this coefficient the contribution ratio of the spin-field interaction is larger that the Dresselhaus interaction, and grows with both the magnetic field and the tilting angle.

\begin{figure}
\onefigure[angle=0,width=80mm]{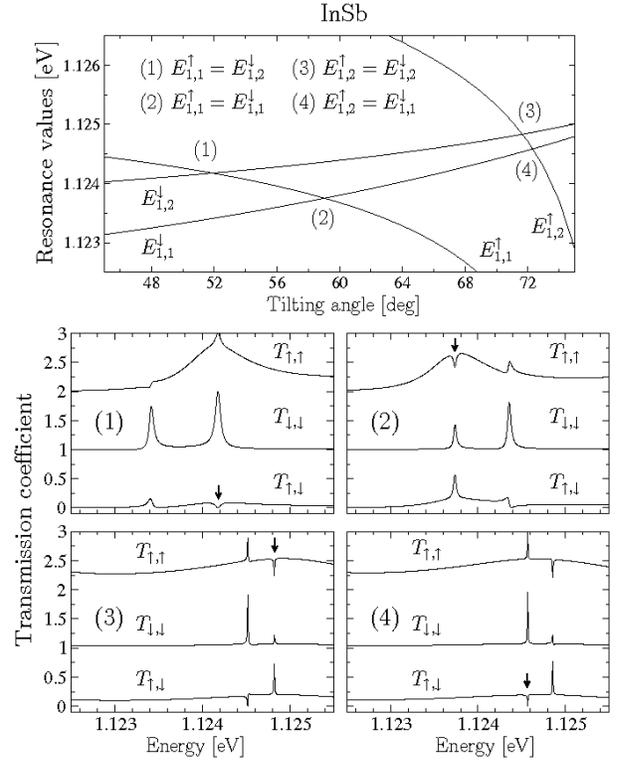}
\caption{ In the top panel, the evolution of the resonant energies $E_{\mu\nu}^{\uparrow}$ and $E_{\mu\nu}^{\downarrow}$ as functions of the tilting angle $\theta_H$ is shown for three-cell $InSb$ HMSL. We plot in the bottom pannels the transmission coefficients for each tilting angle where the resonant energies $E_{\mu\nu}^{\uparrow}$ and $E_{\mu\nu}^{\downarrow}$ cross. Fano anti-resonances appear in $T_{\uparrow, \uparrow}$ (with resonant $T_{\uparrow, \downarrow}$) when $\nu_\uparrow + \nu_\downarrow + n$ is odd (panels (2) and (3)). Fano line-shapes in $T_{\uparrow, \uparrow}$ with vanishing $T_{\uparrow, \downarrow}$ antiresonance appear when the previous sum is even (panels (1) and (4)).}
\label{Fano}
\end{figure}

\begin{figure}
\onefigure[angle=0,width=80mm]{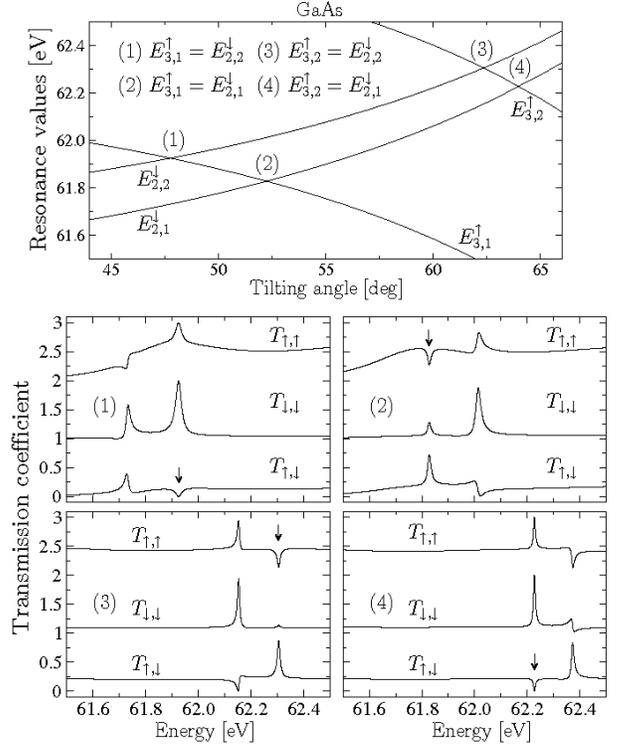}
\caption{ In the top panel, the evolution of the resonant energies $E_{\mu\nu}^{\uparrow}$ and $E_{\mu\nu}^{\downarrow}$ as functions of the tilting angle $\theta_H$ is shown for a three-cell $GaAs$ HMSL. In the bottom pannels, as in fig. \ref{Fano}, we plot the transmission coefficients for each tilting angle where the resonant energies $E_{\mu\nu}^{\uparrow}$ and $E_{\mu\nu}^{\downarrow}$ cross. Fano anti-resonances appear in $T_{\uparrow, \uparrow}$ (with resonant $T_{\uparrow, \downarrow}$) when $\nu_\uparrow + \nu_\downarrow + n$ is odd (panels (2) and (3)). Fano line-shapes in $T_{\uparrow, \uparrow}$ with vanishing $T_{\uparrow, \downarrow}$ antiresonance appear when the previous sum is even (panels (1) and (4)).}
\label{FanoGaAs}
\end{figure}

In ref. [\cite{JLCardoso2}] the band structure of the transmission coefficients $T_{\uparrow, \uparrow}$, $T_{\uparrow, \downarrow}$ and $T_{\downarrow, \downarrow}$ was studied as a function of energy and $\theta_{H}$. It was shown that when increasing $\theta_H$, the band structure of $T_{\uparrow, \uparrow}$ shifts towards lower energies while that of $T_{\downarrow, \downarrow}$ goes to towards higher energies. On the other hand, in ref. [\cite{JLCardoso}], it was also shown that for increasing the magnetic field strength $H$ the transmission coefficient $T_{\uparrow, \uparrow}$ experiences a red shift, while $T_{\downarrow,\downarrow}$ undergoes a blue shift. In fig. \ref{Fano} we show the energy levels and the transmission coefficients of a three-cell superlattice. In the upper frame the evolution of the energy levels as function of the magnetic-field tilting angle $\theta_{H}$ is shown. In these plots $E_{\mu,\nu}^i$ refers to the $\nu$-th resonance in the $\mu$-th subband, and {\it i} denotes the spin component. Though no direct anti-crossing phenomena of the resonant energy levels (characteristic of the perturbation approaches) is observed, Fano-like anti-resonances, leading to strong spin-transitions with well defined selection rules, are systematically found in the transmission coefficients behavior. This interesting spin-field interference phenomenon is behind the sudden increase or decrease of the spin-flip processes. In the lower part of fig. \ref{Fano}, we plot the transmission coefficients for each of the four interception points (1-4) shown in the upper part. In this plots the Fano-like anti-resonances are pointed out with arrows. We found that $T_{\downarrow, \uparrow}$ anti-resonances occur when $\nu_\uparrow + \nu_\downarrow + n$ is even, while $T_{\uparrow, \uparrow}$ anti-resonances occur when the previous sum is odd \footnote{A detailed analysis of these selection rules will be published elsewhere}. A similar behavior is seen in fig. \ref{FanoGaAs}, where instead of an $InSb$ magnetic superlattice, we have a $GaAs$ MSL. The fulfillment of the same selection rules and similar Fano anti-resonances are observed for each of the energy-level crossings shown in the upper part of fig. \ref{FanoGaAs}. Based on these selection rules and the band shift induced by the magnetic field, we propose a two-step spin inversion device consisting of two HMSL's. In general, we shall assume that the incoming electron beam is polarized with spin ${\downarrow}$. The first SL is devoted to invert the largest fraction of these spin $\downarrow$ electrons, while the second SL filters the remaining spin $\downarrow$ flux, so that at the right hand side of the device only spin $\uparrow$ electrons are found, which means $\left| \varphi_\downarrow (2L) \right|^{2} = 0$.
\begin{figure}
\onefigure[angle=0,width=80mm]{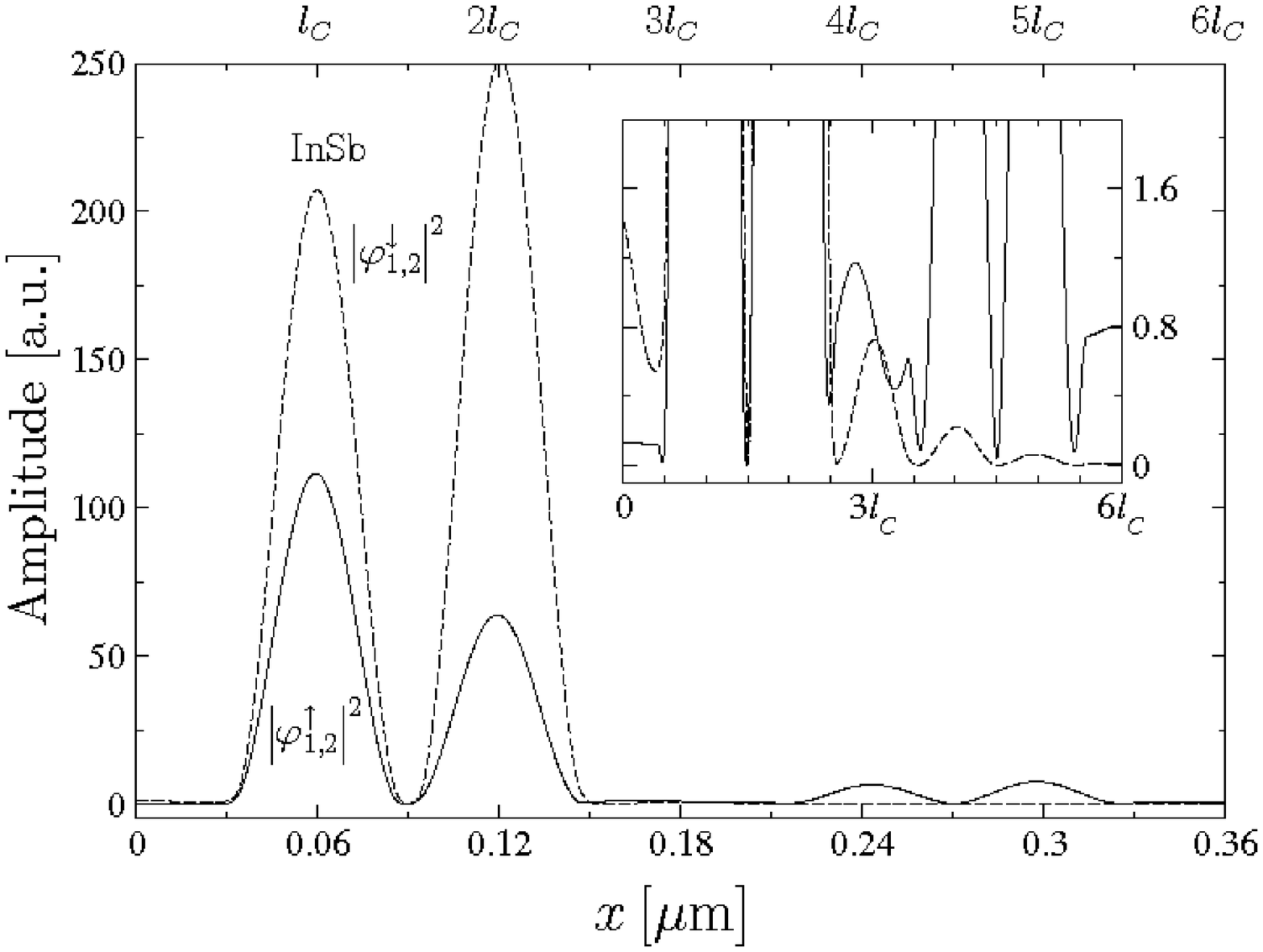}
\caption{ Spin-up $\left| \varphi_\uparrow \right|^{2}$ and spin-down $\left| \varphi_\downarrow \right|^{2}$ wave functions in a two step inversion process based on $InSb$ HMSL's. For a polarized spin $\downarrow$ beam and hight $T_{\uparrow, \downarrow}$ ($\sim 80\%$), the incoming plus the reflected wave function are mainly spin $\uparrow$ at the left, while at the right hand side of the first SL the spin $\uparrow$ flux is higher, with almost null spin$\downarrow$ flux. To visualize better the spin $\uparrow$ and $\downarrow$ electrons distribution with energy $1.1248$eV , the inset shows an enlargement of the distributions.}
\label{inversor}
\end{figure}

The transmission coefficients behavior at resonance and antiresonance energies provides a clear insight on the effects and physical conditions required for the expected device performance. In panels (1) and (4) of figs. \ref{Fano} and \ref{FanoGaAs}, the transition probabilities $T_{\downarrow, \uparrow}$ and $T_{\uparrow, \downarrow}$ vanish at the resonance, while $T_{\uparrow, \uparrow}$ and $T_{\downarrow, \downarrow}$ are practically 1. Under these conditions, the system can neither filter nor polarize the spin fluxes. On the other hand, the anti-resonances of $T_{\uparrow, \uparrow}$ shown in panels (2) and (3) lead to maximum transition probabilities ($T_{\uparrow, \downarrow}$ and $T_{\downarrow, \uparrow}$) with a significant reduction of $T_{\downarrow, \downarrow}$. In these cases one can figure out the possibility of designing a spin-polarizer or a spin-inverter device. Let us now consider two examples of spin-inversion devices, having exactly the same geometrical parameters, but based on different semiconductor medium. One of these devices will be based on $InSb$ and the other one on $GaAs$. The geometrical parameters of each of the three-cell SL's, used in these examples, are: $l_F = w_y = 50$nm and $l_H = 10$nm. In each case the physical conditions corresponding to the first-step SL, are chosen according with panels (3) in figs. \ref{Fano} and \ref{FanoGaAs}.
\begin{figure}
\onefigure[angle=0,width=80mm]{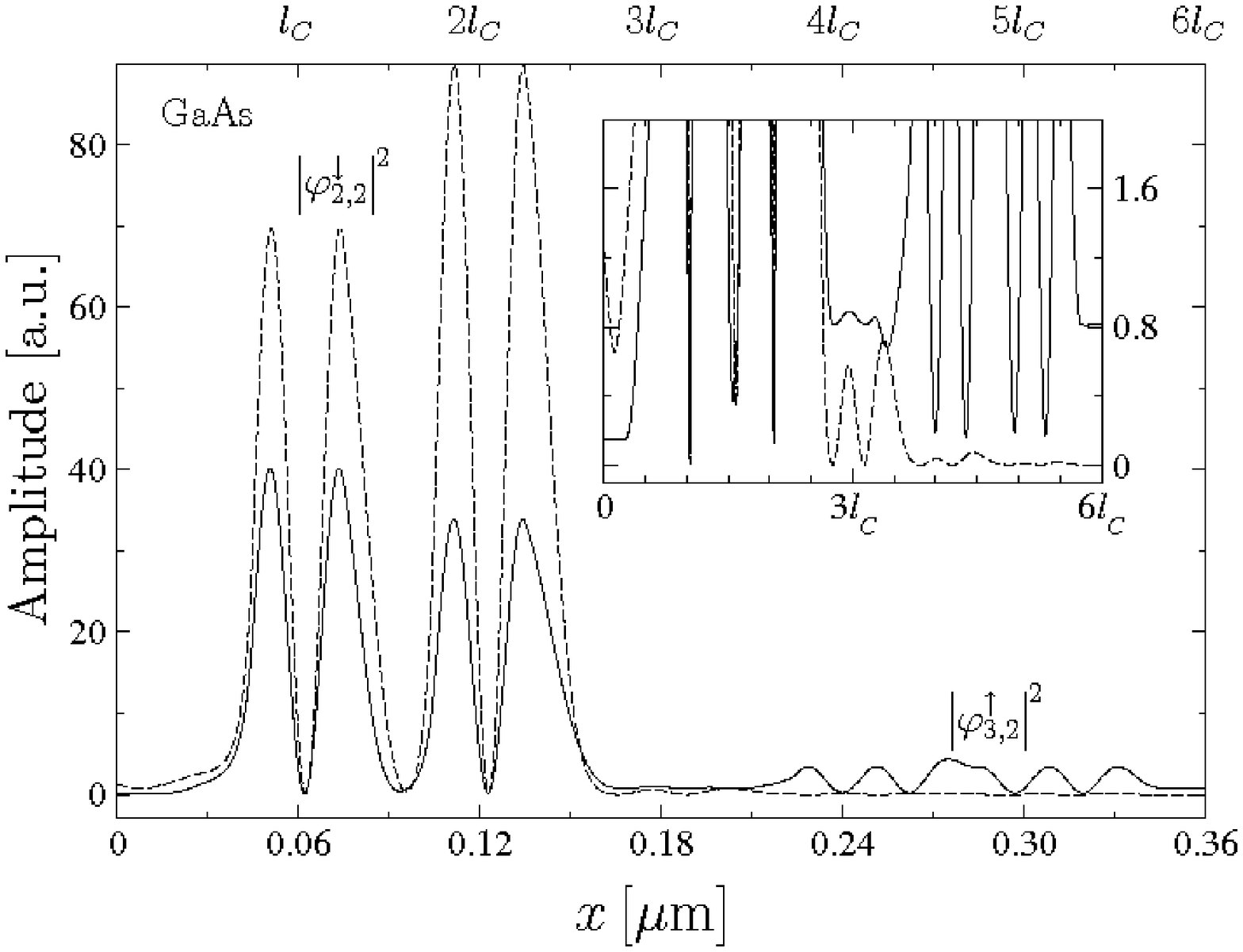}
\caption{ Spin-up $\left| \varphi_\uparrow \right|^{2}$ and spin-down $\left| \varphi_\downarrow \right|^{2}$ wave functions in a two step inversion process. For a polarized spin $\downarrow$ beam and hight $T_{\uparrow, \downarrow}$ ($\sim 80\%$), the incoming plus the reflected wave function are mainly spin $\uparrow$ at the left, while at the right hand side of the first SL the spin $\uparrow$ flux is higher, with almost null spin$\downarrow$ flux. To visualize better the spin $\uparrow$ and $\downarrow$ electrons distribution with energy  $62.3$meV, we amplify these functions in the inset.}
\label{inversorGaAs}
\end{figure}
\begin{figure}
\onefigure[angle=0,width=80mm]{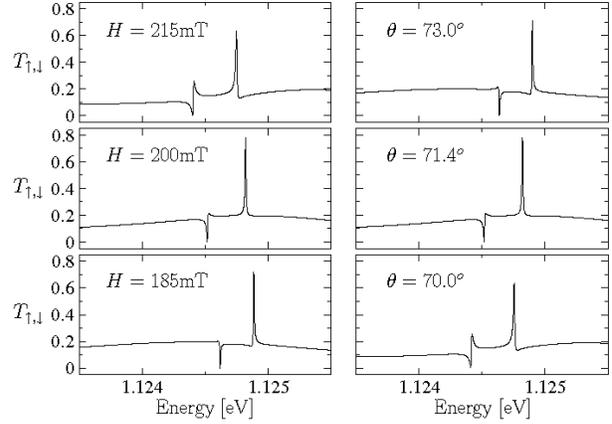}
\caption{The effect of the magnetic field (strength and orientation) uncertaintly on the magnitude and position of the Fano-like resonances of $T_{\uparrow,\downarrow}$. Variations of the order of $1-2 \%$ in $H$ and $\theta$ imply variations of the order of $0.1$meV in the resonances position with practically no change in the magnitude of $T_{\uparrow,\downarrow}$.}
\label{robus}
\end{figure}
In the case of $InSb$ magnetic SL shown in fig. \ref{Fano}, the evolution lines of $E_{1,2}^{\uparrow}$ and $E_{1,2}^{\downarrow}$ at $H=200$mT cross each other when the tilting angle is $\theta _H= 71.44^o$, while for the $GaAs$ SL, shown in fig. \ref{FanoGaAs}, the evolution lines of $E_{3,2}^{\uparrow}$ and $E_{2,2}^{\downarrow}$ at $H=1.5$T cross each other when $\theta _H= 62.4^o$. In the first case, the transition coefficient grows up to $T_{\uparrow, \downarrow}\simeq 0.77$, while for the $GaAs$-spin-inverter $T_{\uparrow, \downarrow}\simeq 0.8$. As mention before, we can visualize the device performance in terms of the carriers probability distributions $\left| \varphi_{\mu,\nu}^\uparrow (x) \right|^{2}$ and $\left| \varphi_{\mu,\nu}^\downarrow (x) \right|^{2}$. In the main graphs of figs. \ref{inversor} and \ref{inversorGaAs} the wave functions $\varphi_{\mu,2}^\uparrow $ (with $\mu$ equal to $1$ and $3$ respectively) follow the behavior of $\varphi_{\mu',2}^\downarrow $, which corroborates the previous discussion. In the presence of large transition amplitude, the spin ${\downarrow}$ beam feeds the spin ${\uparrow}$ electrons population. It is interesting to notice that the spin-flip process preserves the phase information. In a closer view of the wave functions (see insets), the larger fraction of spin ${\uparrow}$ electrons at the r.h.s. of the first steps manifest itself in a larger wave function amplitude ($\varphi_{\mu,2}^\uparrow (3\ell c) \gg \varphi_{\mu,2}^\uparrow (0)$). Notice that even though the incoming spin ${\uparrow}$ wave function is zero, the amplitude at the left of the superlattice is different from zero (note that the incoming electron beam is spin $\downarrow$ polarized), this is due basically to those spin-inverted electrons that are afterwards reflected inside the SL. At variance with this, the spin ${\downarrow}$ amplitude $\varphi_{\mu,2}^\downarrow $ is larger at the left than at the right. This wave function amplitude decreases even more in the second step where it is, finally, completely filtered. For this to happen, the magnetic field has to be tuned in such a way that $T_{\downarrow, \downarrow}\simeq 0$. In our examples, the spin $\downarrow$ filter condition, for ${\theta_H=0}$, is fulfilled when $H=250$mT for $InSb$ and $H=1.53$T for $GaAs$ magnetic superlattices. A small fraction of the converted spin ${\uparrow}$ electrons leave the device at the left. At the end of the second step superlattice, the population of spin $\uparrow$ electrons increases proportionally to $T_{\uparrow, \downarrow}$, up to $\sim 80\%$, while the spin $\downarrow$ population has practically come down to zero. Because of the Dresselhaus interaction a small fraction of the remaining spin $\downarrow$ electrons become spin $\uparrow$, increasing slightly more the spin ${\uparrow}$ flux.

Besides the previous analysis, there are two important issues that deserve some attention. One is the proximity of resonances that could spoil the device performance and the other one is the robustness of these devices against temperature, experimental uncertainties and disorder-induced fluctuations. Concerning the resonances' proximity with each other it is worth mentioning that those electrons whose energy falls inside the neighbor's resonance are completely filtered in the second step that has an operation-energy window (in the case of the $InSb$) of the order of $15$meV. This window implies also relatively low thermal energies hence low operational temperatures, at least with this device. As can be seen from fig. \ref{robus} the robustness against angle and field uncertainties is satisfactory. In this figure we show that the Fano-resonances experience a displacement of the order of $0.1$meV under angle and field strength variations of the order of $1-2\%$. Similar effects appear because of experimental uncertainties with the geometrical parameters.

In this paper we have applied the magnetic superlattice theory \cite{JLCardoso}, in particular the resonant behavior of the spin $\uparrow$ and spin $\downarrow$ transmission coefficents, for designing a device that by using conveniently the Fano anti-resonance selection rules changes the spin projection of almost the whole incident beam. Adding a second filtering step the flux of the outgoing particles becomes fully polarized.

\acknowledgments
We acknowledge A. Robledo-Mart\'{\i}nez, J. Grabinsky, M. Ruggero and M.-W. Wu for useful comments.


\begin{thebibliography}{0}

\bibitem{IZutic}
  \Name{{\u Zuti\'c} I., Fabian J. \and Das Sharma S.}
  \REVIEW{Rev. Mod. Phys.}{76}{2004}{323}

\bibitem{PRHammar}
  \Name{Hammar P. R., Bennet B. R., Yang  M. J.\and Johnson M.}
  \REVIEW{J. Appl. Phys.}{87}{2000}{4665}

\bibitem{GAPrinz} 
  \Name{Prinz G.}
  \REVIEW{Physics Today}{48}{1995}{58}

\bibitem{Fiederling}
  \Name{Fiederling R., Keim M., Reuscher G., Ossau W., Schmidt G., Waag A. \and  Molenkamp L.~W.}
  \REVIEW{Nature}{402}{1999}{787}

\bibitem{Ohno}
  \Name{Ohno Y., Young D. K., Beschoten B., Matsukura F., Ohno H. \and Awschalom D. D.}
  \REVIEW{Nature}{402}{1999}{790}

\bibitem{TKoga}
  \Name{Koga T., Nitta J., Takayanagi H. \and Datta S.}
  \REVIEW{Phys. Rev. Lett.}{88}{2002}{126601}

\bibitem{MKohda}
  \Name{Kohda M., Ohno Y., Takamura K., Matsukura F. \and Ohno H.}
  \REVIEW{Jpn. J. Appl. Phys., Part 1}{40}{2001}{L1274}

\bibitem{EJohnston}
  \Name{Johnston-Halperin E., Lofgreen D., Kawakami R.,  Young D., Coldren L.,  Gossard A. \and Awschalom D.}
  \REVIEW{Phys. Rev. B}{65}{2002}{041306}

\bibitem{YKato}
  \Name{ Kato Y., Myers R. C., Driscoll D. C., Gossard, Levy  J. \and Awschalom D. D.}
  \REVIEW{Science}{299}{2003}{1201}

\bibitem{JMKikkawa1}
  \Name{ Kikkawa J. M., Smorchkova I. P., Samarth N. \and Awschalom D. D.} \REVIEW{Science}{277}{1997}{1284}

\bibitem{JMKikkawa2}
  \Name{ Kikkawa  J. M.\and Awschalom D. D.}
 \REVIEW{Nature}{397}{1999}{139}

\bibitem{JFSong}
  \Name{ Song J. F., Ochiai Y. \and Bird J. P.}
  \REVIEW{Appl. Phys. Lett.}{82}{2003}{4561}

\bibitem{METorio}
  \Name{Torio M. E., Hallberg K., Flach S., Miroshnichenko A. E. \and Titov M.}
  \REVIEW{Eur. Phys. J. B}{37}{2004}{399}

\bibitem{Datta}
  \Name{ Datta S. \and Das B.}
  \REVIEW{Appl. Phys. Lett.}{56}{1990}{665}

\bibitem{Barthelemy}
  \Name{ Barthelemy A. \and Fert A.}
  \REVIEW{Phys. Rev. B}{42}{1991}{13124}

\bibitem{Johnson}
  \Name{ Johnson B.L. \and Camley R.E.}
  \REVIEW{Phys. Rev. B}{44}{1991}{9997}

\bibitem{Berger}
  \Name{ Berger L.}
  \REVIEW{Phys. Rev. B}{54}{1996}{9353}

\bibitem{JLCardoso}
  The main lines of our theoretical approach were introduced in:
  \Name{ Cardoso J. L., Pereyra P. \and Anzaldo-Meneses A.}
  \REVIEW{Phys. Rev. B}{63}{2001}{153301}

\bibitem{Wu}
  \Name{ Wu M. W., Zhou J. \and Shi Q. W.}
  \REVIEW{ Appl. Phys. Lett.}{85}{2004}{1012}

\bibitem{JLCardoso2}
  \Name{ Pereyra P. \and Cardoso J. L.}
  \REVIEW{Phys. Stat. Sol. (c)}{4}{2007}{462}

\bibitem{PPereyra1}
  \Name{ P. A. Mello, P. Pereyra \and N. Kumar}
  \REVIEW{Ann. Phys. (N.Y.)}{181}{1998}{290};
  \Name{ P. Pereyra}
  \REVIEW{ J. Math. Phys. (N.Y.)}{36}{1995}{1166}

\end{thebibliography}
\end{document}